\documentclass[prl,10pt,aps,floatfix,twocolumn,showpacs,preprintnumbers,
  superscriptaddress]{revtex4}
\usepackage{graphicx}
\usepackage{dcolumn}
\usepackage{bm}
\usepackage{amssymb}
\usepackage{pifont}
\usepackage{amsmath}
\usepackage{times}
\usepackage[nooneline]{subfigure}
\usepackage{color}

\begin{document}
\newcommand {\be}{\begin{equation}}
\newcommand {\ee}{\end{equation}}
\newcommand {\ba}{\begin{eqnarray}}
\newcommand {\ea}{\end{eqnarray}}

\title{Covariant hydrodynamic Lyapunov modes and strong 
stochasticity threshold in Hamiltonian lattices}

\author{M. Romero-Bastida}
\email{mromerob@ipn.mx}
\affiliation{SEPI ESIME-Culhuac\'an, Instituto Polit\'ecnico Nacional,
Distrito Federal 04430, Mexico}

\author{Diego Paz\'o}
\email{pazo@ifca.unican.es}
\author{Juan M.\ L{\'o}pez}
\email{lopez@ifca.unican.es}
\affiliation{Instituto de F\'{\i}sica de Cantabria (IFCA), CSIC--UC, E-39005
Santander, Spain}

\date{\today}

\begin{abstract}
We scrutinize the reliability of covariant and Gram-Schmidt Lyapunov vectors 
for capturing hydrodynamic Lyapunov modes (HLMs) in one-dimensional Hamiltonian lattices.
We show that,
in contrast with previous claims, HLMs do exist for any energy density, so that
strong chaos is not essential for the appearance of genuine (covariant) HLMs. 
In contrast, Gram-Schmidt Lyapunov vectors lead to misleading results concerning the
existence of HLMs in the case of weak chaos.
\end{abstract}


\maketitle

\emph{Introduction.-} The existence of hydrodynamic Lyapunov modes (HLMs) in
spatially extended dynamical systems has recently attracted a great deal of
attention. HLMs are collective long-wavelength perturbations associated with the
smallest positive Lyapunov exponents (LEs). They may appear in hard-core and
soft-core potential
systems~\cite{posch00,eckmann00,mcnamara01b,mcnamara04,wijn04,forster05,
tanigu05,morriss09,chung11}, as well as in coupled-map
lattices~\cite{CMLHLM1,CMLHLM2,GoodHLM}, of either Hamiltonian or dissipative
type. HLMs are believed to be connected to a system's macroscopic properties
and to encode valuable information for understandiing universal features of
high-dimensional nonlinear systems~\cite{posch00,tanigu05,wijn04}. Nowadays, we
lack a complete understanding of how or where these modes may show up, although,
it is generally believed that conservation laws and translational invariance are
essential for HLMs to
appear~\cite{mcnamara01,mcnamara01b,wijn04,CMLHLM1,CMLHLM2,GoodHLM}. In the case
of Hamiltonian lattices it has also been argued~\cite{FPUHLM,XYHLM} that the
system must be well above the so-called strong stochasticity threshold
(SST)~\cite{kantz89,Pettini2,Pettini} in order to show significant HLMs. This
stochasticity threshold separates weak and strong dynamical chaos: Below the SST
(low energy density) the relaxation time to energy equipartition grows as a
stretched exponential as the energy density
decreases~\cite{kantz89,Pettini2,Pettini}, whereas above the SST (for high
energy densities) the relaxation time needed to reach energy equipartition is
independent of the energy density. Now, for the Fermi-Pasta-Ulam
(FPU)~\cite{fpu,pettini05} and other Hamiltonian lattice models, it has been
claimed~\cite{FPUHLM,XYHLM} that strong chaos is essential for the existence of
significant HLMs. This relation could suggest a connection between the HLMs and
the ergodic problem, which is closely related to the dynamical foundations of statistical mechanics
in high-dimensional nonintegrable Hamiltonian systems. However, this is indeed a
puzzling situation since, in principle, no matter whether or not energy equipartition
is reached within the observation time, the system is still chaotic with a
finite density of non-zero LEs.

In this work we solve this question by demonstrating that \emph{covariant} HLMs
associated with nearly zero LEs do actually exist in high-dimensional
nonintegrable Hamiltonian systems for \emph{any} energy density. We also show
that the significance measures of these covariant modes are completely
equivalent both above and below the SST. In contrast, if non covariant
Gram-Schmidt (GS) vectors are used, as in previous studies, the existence
of HLMs and their significance do depend on the scalar product convention,
resulting in artifacts that are not intrinsic to the system under study.

Most studies in the existing literature dealing with the problem of computing
HLMs relied on the backward (also called GS) Lyapunov vectors (BLVs) to obtain
the phase-space directions that correspond to the smallest positive LEs. BLVs
are obtained by the standard GS orthonormalization procedure to obtain the
Lyapunov spectrum~\cite{ershov98,benettin80}. BLVs are known to have many
important issues and artifacts~\cite{szendro07,pazo08} that ultimately render
them useless for many purposes. First and foremost, they are forced to form an
orthogonal set; this is not a minor point because it leads to vectors that are
not covariant with the dynamics. Therefore, when left to evolve freely with the
tangent space dynamics, they will not map to themselves but will (exponentially)
rapidly collapse in the direction of the main Lyapunov vector (LV). An even more
important issue for our purposes is the fact that BLVs depend explicitly on the
scalar product needed to define mutual orthogonalization. 

We employ here the so-called \emph{covariant} (or
characteristic~\cite{legras96}) Lyapunov vectors (CLVs)~\cite{eckmann}. This set
of vectors reflects the bona-fide covariant tangent space directions and
provides a genuine Oseledec splitting of phase space (one direction is
unambiguously associated with each nondegenerate LE).
A comparison of the significance of
HLMs obtained from BLVs and CLVs in coupled map lattices
was recently made by Yang and Radons~\cite{yang10},
but it did not reveal important differences. In contrast, we show here that
strong differences appear between BLVs and CLVs below the SST.

\emph{Models.-} The reference Hamiltonian for the one-dimensional lattice models
we are considering can be written as ${\cal H}=\sum_{i=1}^N\left[p_i^2/(2m_i) +
{\cal V}(q_{i+1}-q_i)\right]$, where $N$ is the system size and ${\cal V}(x)$ is
the nearest-neighbor interaction potential. $\{m_i,q_i,p_i\}_{i=1}^N$ are the
dimensionless mass, displacement, and momentum of the $i$th particle, each of
unit mass $m_i=1$; periodic boundary conditions are assumed
($q_{_{N+1}}=q_{_1}$). The models hereafter considered are (i) the FPU
$\beta$--model, with potential ${\cal V}(x) = x^2/2+x^4/4$, and (ii) the $XY$
model with ${\cal V}(x)=1-\cos(x)$. The energy density $\varepsilon=E/N$,
$E$ being the total energy, is the control parameter for the system dynamics. The
integration of the $2N$ equations of motion as well as the computation of the
associated CLVs are performed employing a computationally efficient numerical
algorithm we presented in Ref.~\cite{romero10}. 
In particular, CLVs
are obtained from the intersection of subspaces embedded by GS (backward) and
by so-called forward Lyapunov vectors, according to the formulas by Wolfe and Samelson~\cite{wolfe07}
(see also \cite{pazo08}).
Our numerical implementation
allows us to explore the
low-energy region wherein exceedingly long transients are required. 

\emph{Analysis.-} The significance of HLMs is usually analyzed through the
\emph{Lyapunov vector fluctuation density}, which is a dynamical quantity
defined as $u^{(\alpha)}(q,t)=\sum_i\delta q^{(\alpha)}_i(t)\delta(q-q_i)$,
where $q_i$ is the position coordinate of the $i$th particle and $\delta
q^{(\alpha)}_i$ is the coordinate component of the $\alpha$th LV
$\left[\delta\mathbf{q}^{\left(\alpha\right)}\left(t\right);
\delta\mathbf{p}^{\left(\alpha\right)}\left(t\right)\right]$ for the $i$th
particle. In order to detect the existence of wavelike or spatially extended
modes, we apply a Fourier transform, $\hat u^{(\alpha)}(k,t)=\int
(dq/2\pi)\exp(\mathrm{i}kq)u^{(\alpha)}(q,t)$, and then compute the stationary
structure factor of $u^{(\alpha)}(q,t)$,
$S^{(\alpha)}(k)=\lim_{t\rightarrow\infty}\langle\hat u^{(\alpha)}(k,t)\hat
u^{(\alpha)}(-k,t)\rangle$, where $\langle\cdots\rangle$ stands for time
average. In order to compare the relative weight of the spectrum maximum with
respect to the background power we normalize by the area below the curve and
define $\bar{S}^{(\alpha)} (k) \equiv S^{(\alpha)}(k)/\sum_k S^{(\alpha)}(k)$.
Typically the spectra for LVs associated with near-zero positive LEs will
exhibit a maximum at some $k_\mathrm{max}(\alpha)$.

For a quantitative determination of the existence of HLMs, two complementary
measures can be used~\cite{CMLHLM2,FPUHLM,XYHLM,3DLJHLM}:
$\bar{S}^{(\alpha)}(k_{\mathrm{max}})$ and the spectral entropy $H^{(\alpha)}$.
Here $\bar{S}^{(\alpha)}(k_{\mathrm{max}})$ denotes the height of the $\alpha$th
LV stationary structure factor at its maximum, whereas the spectral entropy
$H^{(\alpha)}$ is defined as $H^{(\alpha)} \equiv -\sum_k
\bar{S}^{(\alpha)}(k)\ln \bar{S}^{(\alpha)}(k)$ and measures how the normalized
spectral power $\bar{S}^{(\alpha)}(k)$ spreads among wave numbers $k$ for a given
Lyapunov index $\alpha$. Lower values of the entropy indicate that most of the
spectral power is concentrated around fewer wavelengths. Therefore, a small
value of $H^{(\alpha)}$ and a corresponding large value of
$\bar{S}^{(\alpha)}(k_{\mathrm{max}})$ at some index $\alpha$ indicate the
existence of a sharp, well-defined peak of the structure factor for the
$\alpha$th LV, indicating that a particular spatial wavelength is favored. HLMs,
if present, would correspond to extended macroscopic modes with a
long wavelength, $k_{\mathrm{max}}(\alpha) \to 0$, associated with the smallest
positive LEs, $\lambda_\alpha \to 0^{+}$.
\begin{figure}
\includegraphics[width=0.99\linewidth,angle=0.0]{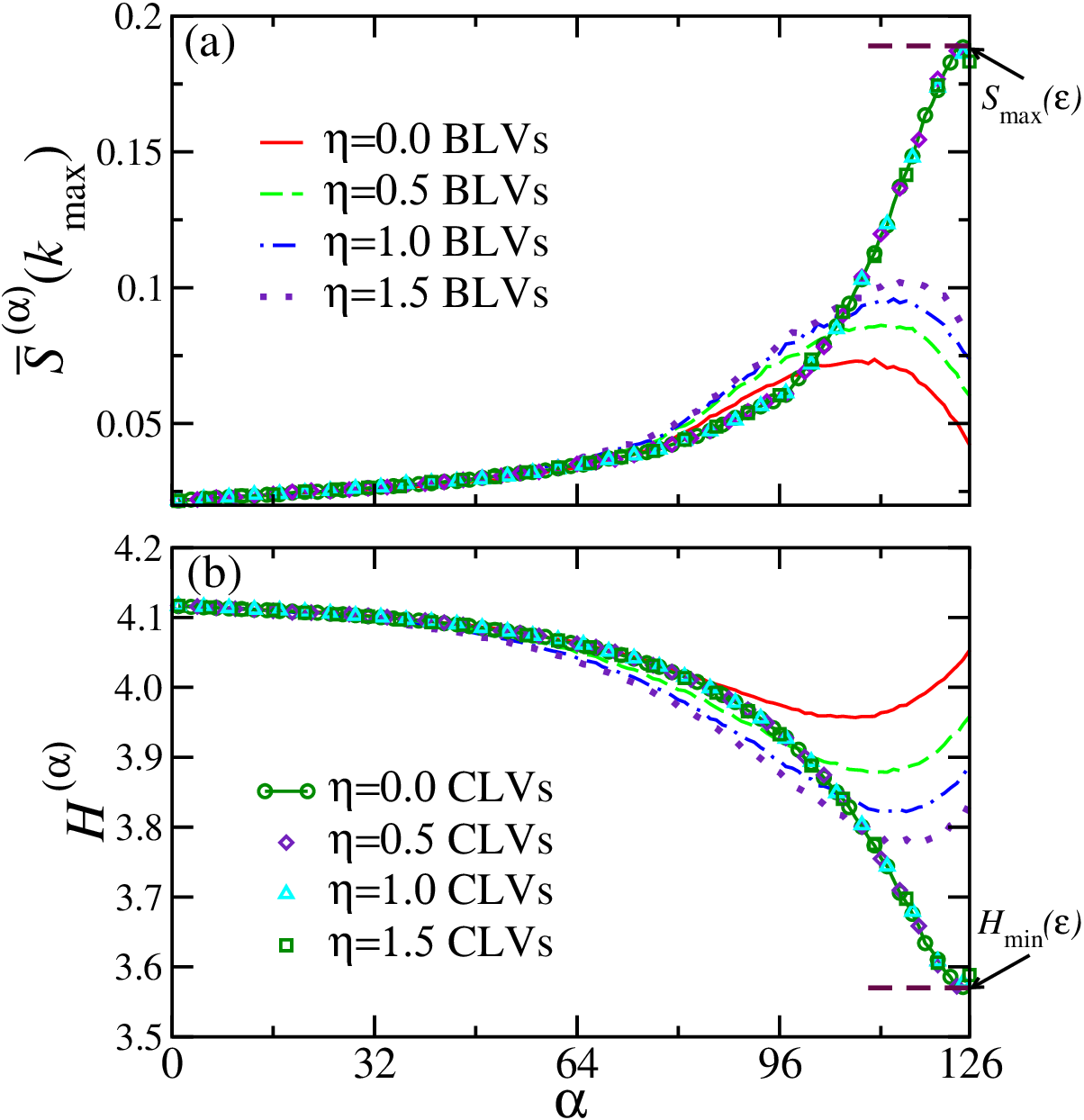}
\caption{(Color online) Above the SST: (a) Normalized
$\bar{S}^{(\alpha)}(k_{\mathrm{max}})$ and (b) spectral entropy $H^{(\alpha)}$
{\it vs.} Lyapunov index $\alpha$ corresponding to the BLVs (lines) and CLVs
(symbols) for different values of the perturbation parameter $\eta$ of the
scalar product (see text). Both plots correspond to the FPU model with $N=128$
and $\varepsilon=10$.} 
\label{fig:SkHs}
\end{figure}

\begin{figure}
\includegraphics[width=0.90\linewidth,angle=0.0]{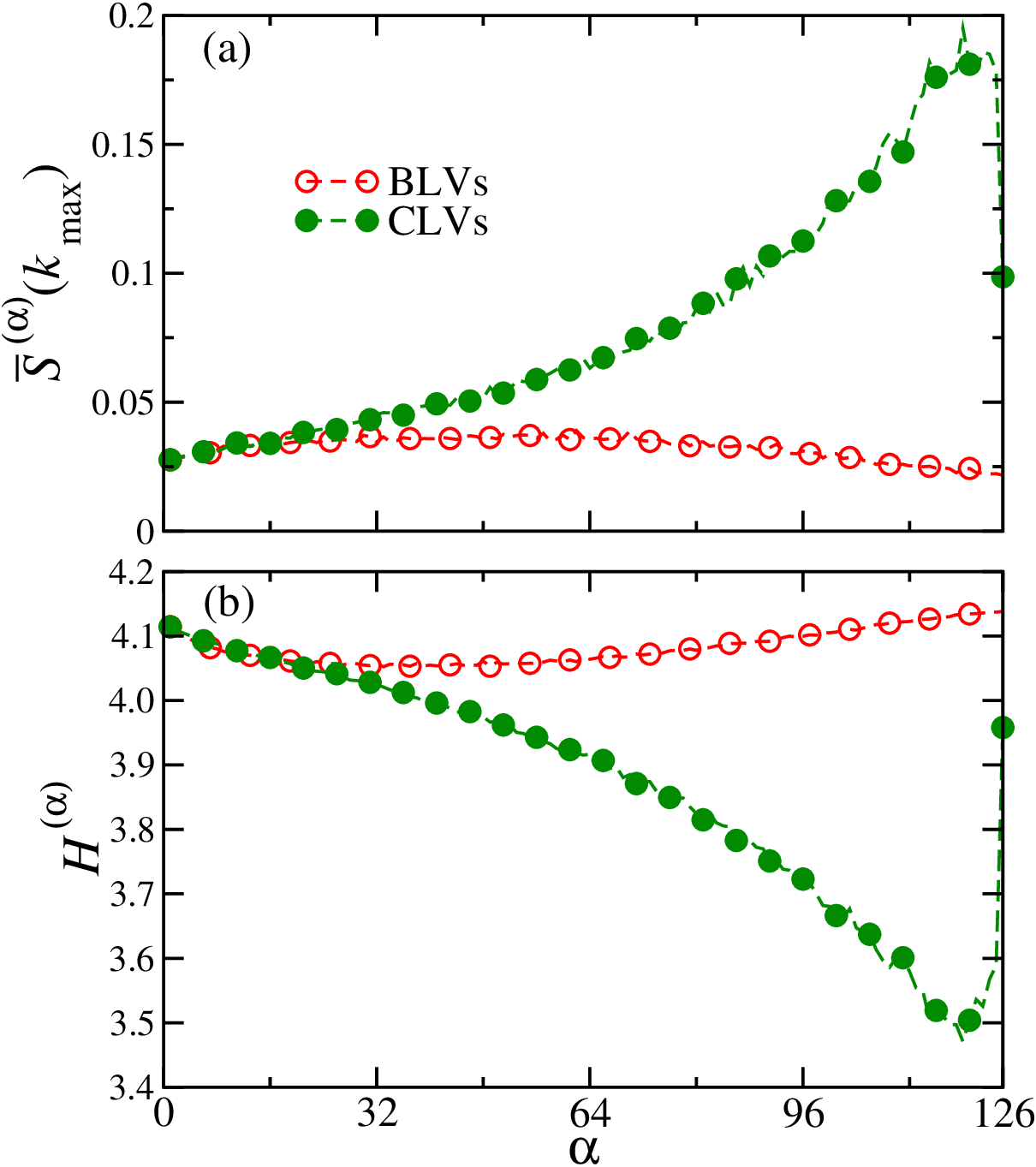}
\caption{(Color online) Below the SST: (a)
$\bar{S}^{(\alpha)}(k_{\mathrm{max}})$ and (b) spectral entropy $H^{(\alpha)}$
{\it vs.} the LV index $\alpha$ for the BLVs (open circles) and CLVs (filled
circles). Both plots correspond to the FPU model with $N=128$, energy density
$\varepsilon=0.01$, and $\eta=0$.}
\label{fig:SkHse0p01}
\end{figure}

\emph{Results.-} Our first result concerns the dependence of the significance
measure of HLMs on the metric used
when constructing the set of backward LVs. 
Since some scalar product always needs to be defined 
(for Gram-Schmidt orthonormalization)
--and this is to a great extent an arbitrary choice-- meaningful conclusions can only
be obtained if the significance measures
are independent of the chosen scalar product. In order to emphasize the impact of the scalar product when
searching for HLMs, we use the scalar product,
$\left[\delta\mathbf{q}^{\left(\alpha\right)}\left(t\right);
\delta\mathbf{p}^{\left(\alpha\right)}\left(t\right)\right] \mathbb{M}_{2N\times
2N}\left[\delta\mathbf{q}^{\left(\alpha\right)}\left(t\right);
\delta\mathbf{p}^{\left(\alpha\right)}\left(t\right)\right]^T$, with the matrix
$(\mathbb{M})_{i,j} = \delta_{i,j} + \eta \; \delta_{i \leq N,j \leq N}$,
$\eta$ being an arbitrary constant. This corresponds to the perturbed Euclidean norm
$\sqrt{(\delta\mathbf{q})^2(1+\eta) + (\delta\mathbf{p})^2}$. We computed the
BLVs and CLVs for different choices of $\eta$ and the results for the measures
$\bar{S}^{\alpha}(k_{\mathrm{max}})$ and $H^{(\alpha)}$ for the FPU model in the
strongly chaotic regime for a high energy density $\varepsilon=10$ are presented
in Fig.~\ref{fig:SkHs}. For both sets of vectors it is clear that
$\bar{S}^{\alpha}(k_{\mathrm{max}})$ presents a maximum and $H^{(\alpha)}$ a
minimum for some index $\alpha_\mathrm{max}$. The curve in
Fig.~\ref{fig:SkHs}(a) for the BLV with $\eta = 0$ corresponds to the mode
reported in Ref.~\cite{FPUHLM} as a truly HLM for the FPU system. However, note
that $\alpha_\mathrm{max} \approx 102$ is still far from $126$ by $20 \%$, which
is the Lyapunov index value corresponding to the smallest positive LE. This
deviation was attributed in Ref.~\cite{FPUHLM} to fluctuations of finite-time
LEs. We claim that this mode is not a good HLM. As clearly seen in
Fig.~\ref{fig:SkHs}(a), BLV curves exhibit an $\alpha_\mathrm{max}$ that
progressively shifts as $\eta$ increases, demonstrating that the mode is not
intrinsic and its position strongly depends on the employed scalar product. In
contrast, if CLVs are used instead, the positions of the extrema, as well as
the significance measures $\bar{S}^{(\alpha)}(k_\mathrm{max})$ and
$H^{(\alpha)}$,
are independent of
the $\eta$ value used, as
confirmed by Fig.~\ref{fig:SkHs}. Also, the positions of both extrema are
located at LE index $\alpha_\mathrm{max}\approx126$, exactly where the smallest
positive LE is located. These results indicate that only CLVs can generically
detect the existence of intrinsic (scalar-product independent) HLMs. This
becomes even clearer if we compute the significance measures
$\bar{S}^{(\alpha)}(k_{\mathrm{max}})$ and $H^{(\alpha)}$ for energy densities
well below the SST, which is around $\varepsilon_c \approx 0.20$ for the
FPU~\cite{FPUHLM} (note also an earlier estimate of $\varepsilon_c \approx
0.12$~\cite{casetti93}). In this weakly chaotic regime it has been reported that
HLMs fail to exist~\cite{FPUHLM,XYHLM}. In Fig.~\ref{fig:SkHse0p01} we plot
$\bar{S}^{(\alpha)}(k_{\mathrm{max}})$ and $H^{(\alpha)}$ for an extremely low
energy density $\varepsilon=0.01$, where the system behaves almost as a harmonic
oscillator chain. As can be readily seen, if CLVs are used, the maximum of
$\bar{S}^{(\alpha)}(k_{\mathrm{max}})$ corresponds to the minimum of
$H^{(\alpha)}$ at the Lyapunov index $\alpha_\mathrm{max}\approx 120$. Thus
covariant HLMs exist even well below the SST. In contrast, BLVs fail to detect
any significant HLMs for $\varepsilon = 0.01$ (actually, for any energy density
below $\varepsilon_c$), which again indicates that BLVs are unreliable objects
whereby to detect HLMs. 

\begin{figure}
\includegraphics[width=0.90\linewidth,angle=0.0]{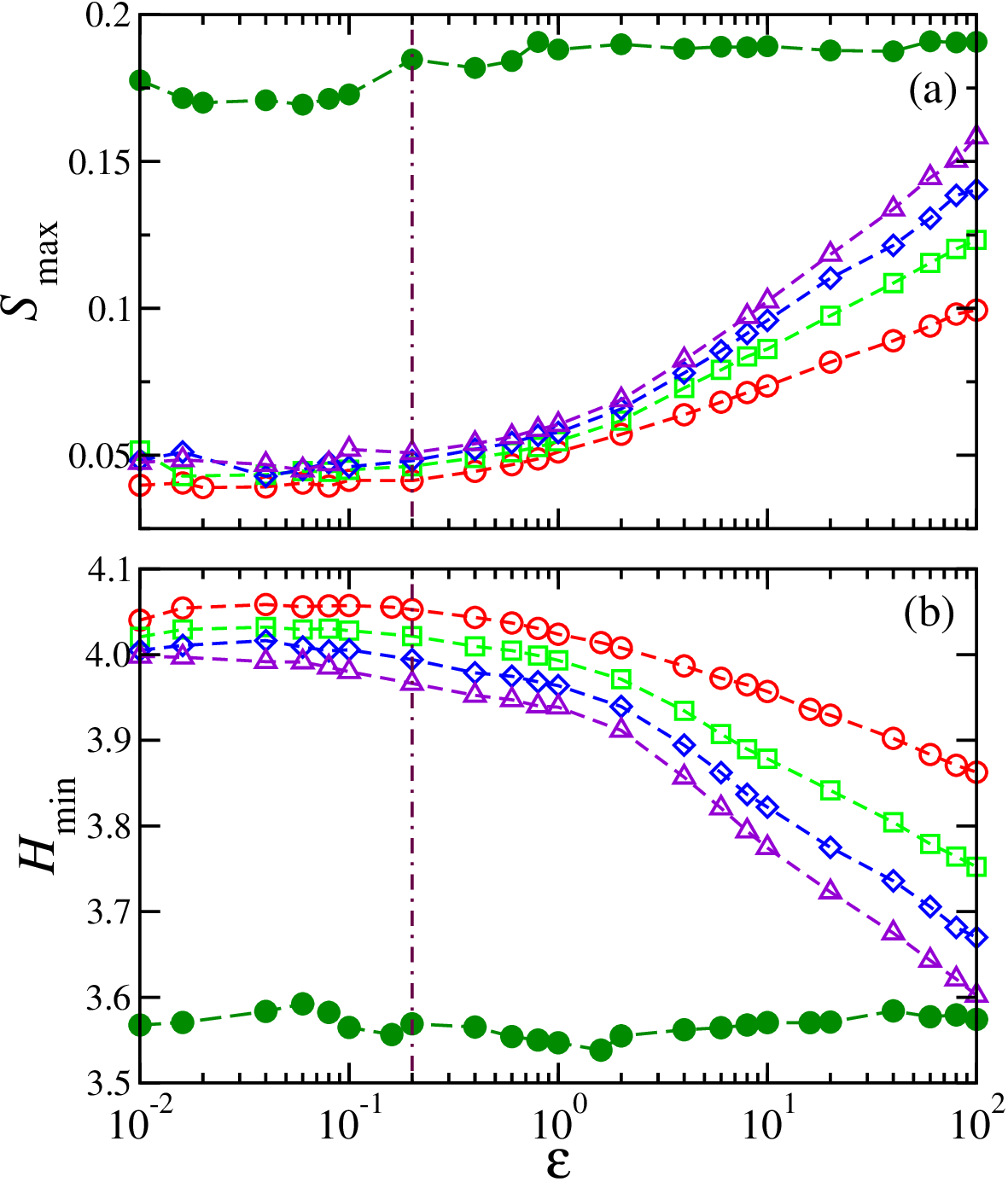} 
\caption{(Color online) (a) $S_{\mathrm{max}}$ and (b) $H_{\mathrm{min}}$ {\it
vs.} energy density $\varepsilon$ for the BLVs using different scalar products
(see text): $\eta=0$ (open circles), $\eta=0.5$ (squares), $\eta=1$, (diamonds)
and $\eta=1.5$ (triangles). Data points computed by means of the CLVs are shown
with filled circles, and vertical dot-dashed lines indicate the position of the SST
according to Ref.~\cite{FPUHLM}.}
\label{fig:SHvseps}
\end{figure}

We can now quantify the significance of the HLMs as a function of the energy
density by plotting in Fig.~\ref{fig:SHvseps} the extreme values
$S_{\mathrm{max}}$ and $H_{\mathrm{min}}$ indicated in Fig.~\ref{fig:SkHs}. The
energy dependence of these extreme values also shows striking differences when
BLVs or CLVs are used. Again, different scalar products lead to different curves
if BLVs are used. GS HLMs seem to appear only above some energy density. Indeed,
a crossover toward an increasing significance of HLMs is observed as the energy
density increases. However, as shown in Fig.~\ref{fig:SHvseps}, the curves for
BLVs display a crossover that shifts toward lower $\varepsilon$ values as
$\eta$ increases. Thus, the relation of this crossover point with the SST
$\varepsilon_c$, first conjectured in Ref.~\cite{FPUHLM}, turns out to be an
artifact that arises from the use of BLVs, since its position explicitly depends
on the employed scalar product. In contrast, the maximum power
$S_{\mathrm{max}}(\varepsilon)$ and minimum spectral entropy
$H_{\mathrm{min}}(\varepsilon)$ remain constant for the full range of energy
densities if CLVs are used to identify HLMs. Figure~\ref{fig:SHvseps} summarizes
our main result: covariant HLMs do exist and their significance, in qualitative
and quantitative terms, is the same for the whole range of energy densities
studied, irrespective of whether we are above or below the SST.

\begin{figure}
\includegraphics[width=0.90\linewidth,angle=0.0]{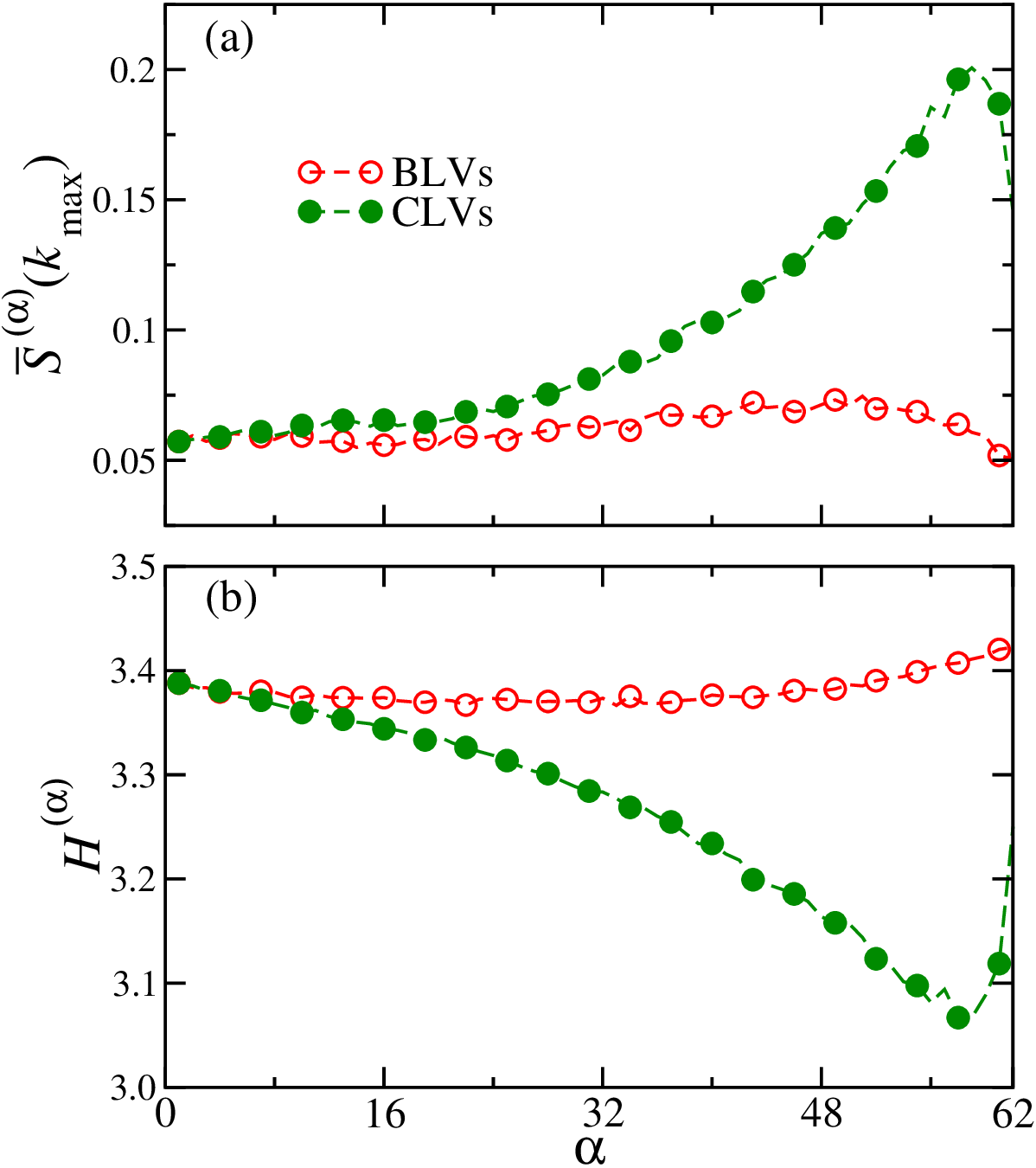}
\caption{(Color online) (a) Maximum power spectrum
$\bar{S}^{(\alpha)}(k_{\mathrm{max}})$ and (b) spectral entropy $H^{(\alpha)}$
for the $XY$ model with $N=64$, $\varepsilon=0.1$, and $\eta=0$, well below the
SST.}
\label{fig:xye0p1}
\end{figure}

Next we briefly discuss our results for the $XY$ model consisting of a
one-dimensional lattice of interacting rotors. For large enough energy
densities, individual particles become nearly independent and the system is
close to a collection of free rotors, whereas for small energy densities the
system is in a near-harmonic regime, analogous to the corresponding one in the
FPU model. We have studied the existence of HLMs in the $XY$ model and obtained
results that are qualitatively similar to those in the FPU case. In the high
energy density regime the interaction potential is not strong enough to maintain
the needed coupling between neighboring rotors; thus spatially extended
perturbations cannot be supported by the system dynamics and no HLMs were
observed in this regime. In the opposite limit of extremely low energy densities
we found no HLMs if BLVs are used, in agreement with an earlier
study~\cite{XYHLM}; however, covariant HLMs do exist even for a low energy
density of $\varepsilon = 0.1$, as can be appreciated in Fig.~\ref{fig:xye0p1}.
The numerical values of the significance measures depend again on the scalar
product if BLVs are used, while no effect is seen for CLVs, as expected from our
discussion of the FPU case. This confirms our finding that strong chaos is {\em
not} essential for the appearance of genuine covariant HLMs.

\emph{Conclusions.-} We have studied HLMs in Hamiltonian lattices and compared
covariant with backward LVs. There are strong theoretical arguments which
suggest that HLMs are associated with the existence of conservation laws and
translational invariance~\cite{mcnamara01,wijn04,CMLHLM2} of the equations of
motion. These {\em exact} symmetries are thus expected to be satisfied
irrespective of the energy density. We have shown that covariant HLMs indeed
exist for any energy density so that strong chaos is not an essential
requirement for the appearance of genuine HLMs. Actually, the significance of
covariant HLMs remains constant both above and below the SST, which seems to
indicate that the mechanisms held responsible of the existence of both phenomena
should be entirely different. We have also demonstrated that BLVs lead to
misleading results concerning the existence of good HLMs and to significance
measures that display an undesired dependence on the scalar product. 

The spatial structure of the covariant HLMs herein studied indicates that these
extended modes have a scale-invariant structure, which contrasts with the
truly wavelike spatially periodic form characteristic of hard-core
systems~\cite{posch00,eckmann00,mcnamara01b,mcnamara04,wijn04,forster05,
tanigu05,morriss09,chung11}. It remains to be understood whether the extended
modes in these two types of systems actually belong to the same kind of
phenomenon, and whether these modes can be related to hydrodynamic properties
for arbitrary systems with extended chaos.

\acknowledgments
M. R. B. thanks CONACyT, Mexico for financial support and Universidad de
Cantabria for partial travel support. We acknowledge financial support from
MICINN (Spain) through Project No. FIS2009-12964-C05-05.

\end{document}